\documentclass[preprint,12pt]{elsarticle}

\usepackage[utf8]{inputenc}
\usepackage[T1]{fontenc}
\usepackage{mathptmx}
\usepackage{amsmath,amssymb}
\usepackage{bm}
\usepackage{graphicx}
\usepackage{tikz}%
\usepackage{xcolor}
\usepackage{hyperref}

\ifdefined\arxivdraft
  \usepackage{setspace}
  \usepackage[mathlines]{lineno}\linenumbers
\fi

\newcommand{\bfv}{\boldsymbol{v}}

\journal{arXiv}

\begin{document}

\begin{frontmatter}

\title{Data-free continuous-time physics-informed neural networks for vortex shedding: a consistent pressure across time windows}

\author[loria]{Micka\"el Delcey}
\author[loria]{Cyril Regan}
\author[kyutech]{Masaki Fuchiwaki}
\author[lemta]{Yoann Cheny\corref{cor1}}
\ead{yoann.cheny@univ-lorraine.fr}
\cortext[cor1]{Corresponding author}

\affiliation[loria]{organization={Universit\'e de Lorraine, CNRS, Inria, LORIA},
            city={F-54000 Nancy},
            country={France}}
\affiliation[lemta]{organization={Universit\'e de Lorraine, CNRS, LEMTA},
            city={F-54000 Nancy},
            country={France}}
\affiliation[kyutech]{organization={Department of Creative Informatics, Kyushu Institute of Technology},
            city={Iizuka},
            country={Japan}}

\begin{abstract}
Without data, physics-informed neural networks (PINNs) in continuous time have so far failed to develop vortex shedding past a circular cylinder in a domain symmetric about the wake centerline: the training converges to the steady symmetric solution. We obtain shedding at $\mathit{Re}=100$ with a continuous-time PINN trained over successive time windows, without any reference data and without temporal discretization, the symmetry being broken by an analytical, divergence-free initial condition. The Strouhal number and the mean drag coefficient agree with an OpenFOAM solution computed on the same domain to within $0.5\%$. The pressure, however, is discontinuous at the junctions between windows, and the lift coefficient is a staircase. We relate this discontinuity to the differential-algebraic structure of the incompressible Navier--Stokes equations, of index~2 after spatial discretization: a consistent initial condition of each window must satisfy the constraint $\nabla\cdot\bfv=0$ for the velocity, and the hidden constraint $\nabla\cdot\partial_t\bfv=0$, with $\partial_t\bfv$ given by the momentum equation, then determines the pressure. Imposing the pressure alone makes the solution continuous but not consistent; our remedy imposes the pressure at the junctions and adds this hidden constraint to the loss. Over five shedding periods, the jumps of the lift coefficient become $20$ times smaller.
\end{abstract}


\end{frontmatter}

\section{Introduction}\label{intro}

Since the work of Raissi et al.~\cite{raissi2019physics}, physics-informed neural networks (PINNs) have been considered as an alternative to classical computational fluid dynamics (CFD) solvers~\cite{cai2022physics,cuomo2022scientific}. A PINN approximates the solution of a partial differential equation by a neural network whose parameters minimize the residuals of the governing equations at collocation points, the derivatives being computed by automatic differentiation~\cite{baydin2018automatic}. The method requires no mesh, and the solver reduces to the evaluation of these residuals within a standard training loop, so that the code is short and easy to maintain. Much effort has been devoted to bringing PINNs to the accuracy of CFD solvers: changes of the network architecture, such as periodic activation functions~\cite{sitzmann2020siren} or Fourier-feature embeddings of the inputs, which counter the difficulty of standard networks in learning high frequencies~\cite{tancik2020fourier}, the balancing of the terms of the loss~\cite{xiang2021self}, the enforcement of temporal causality~\cite{wang2024respecting}, and the combination of such techniques into a training procedure~\cite{wang2023expert}. Problems that have long served as standard benchmarks in CFD nevertheless remain difficult for PINNs. The two-dimensional flow past a circular cylinder~\cite{schafer1996benchmark,williamson1996vortex}, the subject of this work, is an example.

Without data, i.e., for the forward problem, in which the flow is determined by the governing equations and by initial and boundary conditions given in closed form, continuous-time PINNs have so far failed to develop vortex shedding past a circular cylinder in a domain symmetric about the wake centerline. Chuang and Barba~\cite{chuang2022experience,chuang2023predictive} report that the PINN behaves as a steady-flow solver at $\mathit{Re}=200$, and Rohrhofer et al.~\cite{rohrhofer2023role} observe the same behavior at $\mathit{Re}=100$, even when DNS data are supplied over the first half shedding period. Without data, continuous-time PINNs also fail~\cite{wandel2022spline,shah2026complementary} on the Sch\"afer--Turek benchmark~\cite{schafer1996benchmark}, even though the cylinder is placed off the channel centerline~\cite{shah2026complementary}. \c{C}{\i}b{\i}k~\cite{cibik2026} trains a single space-time network with an artificial-compressibility formulation and adds a small antisymmetric transverse perturbation to the initial condition to break the symmetry; the solution nevertheless relaxes to a steady, nearly symmetric wake. The steady symmetric solution cancels the residuals and satisfies the boundary conditions at every instant, so that only the initial-condition term of the loss penalizes it.

Successful simulations of shedding by PINNs therefore supply additional information: reference data in the loss, inside the domain~\cite{shah2026complementary,cibik2026} or as boundary and initial conditions on a subdomain of the wake~\cite{jin2021nsfnets}, an initial condition taken from a numerical simulation~\cite{wang2023expert,khan2026physics}, or a cylinder placed off the channel centerline~\cite{wang2023expert,wei2026simple}.

Other formulations obtain shedding without data by discretizing time. Hu and Senocak~\cite{hu2025unsupervised} do so in a symmetric domain at $\mathit{Re}=100$: the time derivative is discretized by a second-order backward differentiation formula with $\Delta t=0.25$, a separate network representing the flow at each time level, and they motivate this choice by the accumulation of errors observed with space-time formulations. Neural solvers trained without data on a grid also advance the flow from one time step to the next: a recurrent convolutional network on a staggered grid, with a discretized time derivative~\cite{wandel2021learning}, and Spline-PINN~\cite{wandel2022spline}, in which a convolutional network advances the coefficients of Hermite splines in space and time from $\hat t$ to $\hat t+\mathrm{d}t$, on the Sch\"afer--Turek benchmark. Within a time step, the Spline-PINN solution is continuous in time. The approach of the present work uses neither a grid nor a recurrent network: each time window is represented by a network continuous in $(x,y,t)$, trained at collocation points.

Successive time windows~\cite{wight2020solving,krishnapriyan2021characterizing,wang2023expert} keep time as a continuous input of the network: the time interval is divided into windows trained one after the other, the initial condition of each window being the prediction of the previous network at the end of its window. Within a window no temporal discretization is introduced. The junctions between windows are, however, a known weak point. For a rotating cylinder, Khan et al.~\cite{khan2026physics} impose as initial condition of each window the velocity predicted by the previous network, but not the pressure; we call this junction the $\bfv$-junction. They observe a jump of the forces at the boundary between two windows, which they attribute to the time marching not fully enforcing the continuity of the pressure. Their PINN starts from a CFD velocity field in the periodic regime and covers less than one shedding period, two windows in total. Roy and Castonguay~\cite{roy2024exact} enforce the continuity between windows exactly, through a solution ansatz, the required order of continuity being one less than the highest order of the time derivatives in the problem; their examples do not include the incompressible Navier--Stokes equations.

In this work, we make two contributions. First, we show that a continuous-time PINN trained over successive time windows develops vortex shedding at $\mathit{Re}=100$ in a domain symmetric about the wake centerline, without any reference data. The symmetry is broken by an analytical, divergence-free initial condition, two uniform streams of velocities $3/2$ and $1/2$ on either side of the wake centerline, adapted from the initial condition used to trigger shedding with the LS-STAG cut-cell method~\cite{cheny2010ls}. Over 145 windows, the Strouhal number and the mean drag coefficient agree with an OpenFOAM solution computed on the same domain. Second, we show that the force jumps at the junctions come from the pressure, and we relate them to the differential-algebraic structure of the incompressible Navier--Stokes equations, whose spatial discretizations are differential-algebraic systems of index~2~\cite{hairer1989,wan2019accuracy}. A consistent initial condition of such a system must satisfy the constraint $\nabla\cdot\bfv=0$ for the velocity, and the hidden constraint $\nabla\cdot\partial_t\bfv=0$, with $\partial_t\bfv$ given by the momentum equation, then determines the pressure. For such systems, time integrators lose accuracy on the algebraic variable, the pressure~\cite{hairer1989,wan2019accuracy}. By analogy with this loss of accuracy, we interpret the jumps as follows: with the $\bfv$-junction, the initial condition of each window fixes the velocity only, and the loss, which controls $\nabla\cdot\bfv$ in mean square over a short window, determines the pressure at the junction only weakly. The continuity of the velocity, which is the order of continuity prescribed for problems of first order in time~\cite{roy2024exact}, is therefore not sufficient for this differential-algebraic structure. We propose a remedy, the consistent junction, which imposes the pressure in the initial condition of each window and adds this hidden constraint to the loss. Two controls, each with one ingredient only, show that neither suffices alone: the $(\bfv,p)$-junction, which adds the pressure to the initial condition of the $\bfv$-junction but not the hidden constraint to the loss, makes the solution continuous but not consistent, and the $\bfv$-junction with the hidden constraint in the loss reduces the jumps without removing them. The argument does not depend on the geometry and applies to any continuous-time PINN trained over successive windows on the incompressible Navier--Stokes equations; it is tested here on the cylinder only.

Section~\ref{method} describes the problem and the method, and Section~\ref{results} presents the results.

\section{\label{method}Problem and method}

\subsection{Governing equations, domain and initial condition}
With the cylinder diameter $D$ and the free-stream velocity $U_\infty$ as scales, the incompressible Navier--Stokes equations read
\begin{subequations}\label{eq:NS}
\begin{align}
    & \partial_t\bfv + (\bfv\cdot\nabla)\bfv + \nabla p - \frac{1}{\mathit{Re}}\nabla^2\bfv = \boldsymbol{0}, \label{eq:mom}\\
    & \nabla\cdot\bfv = 0, \label{eq:cont}
\end{align}
\end{subequations}
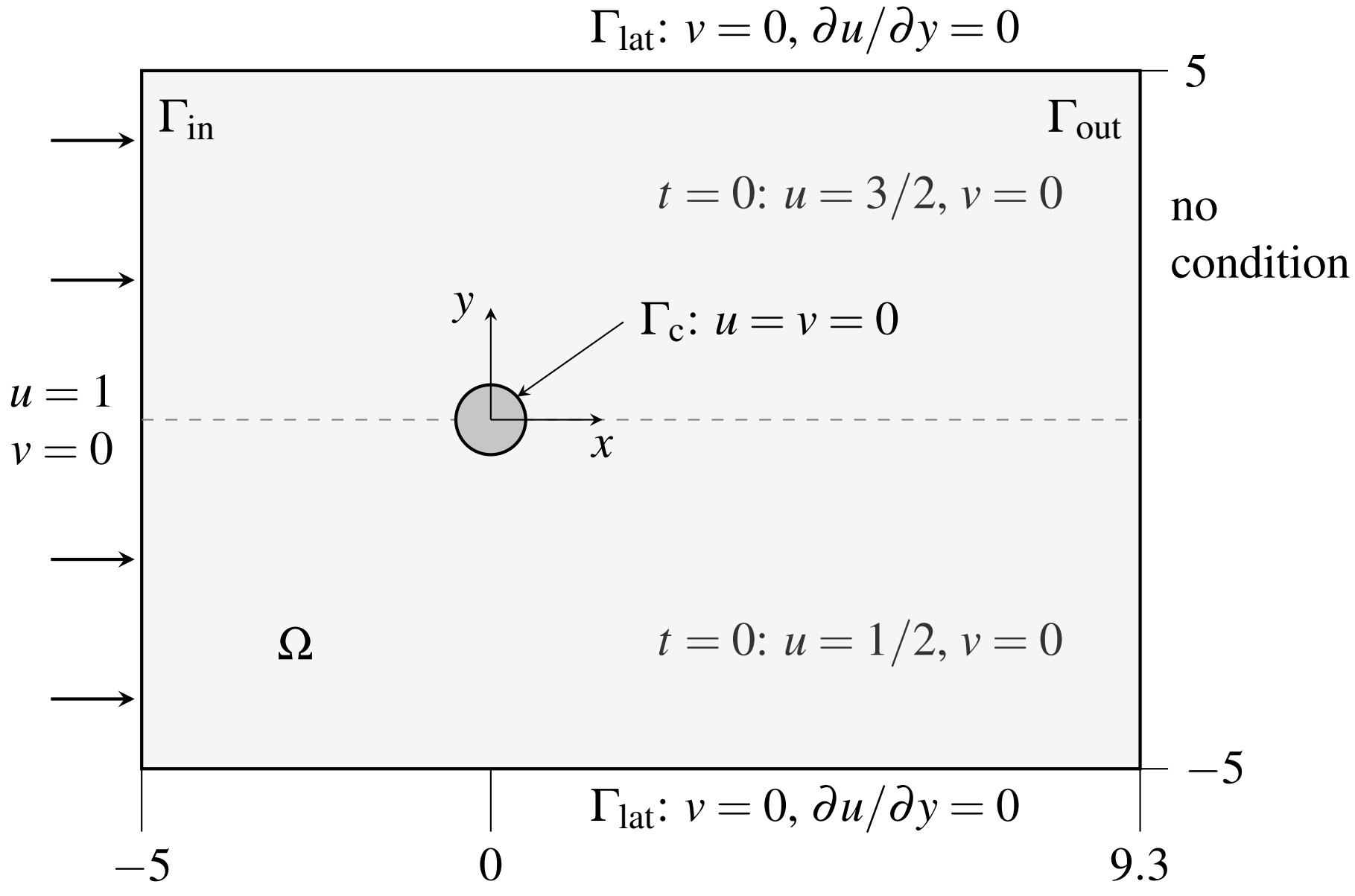
\begin{figure}[ht!]
\centering
\begin{tikzpicture}[scale=0.62, >=stealth]
    \def\xin{-5}\def\xout{9.3}\def\H{5}
    \fill[gray!8] (\xin,-\H) rectangle (\xout,\H);
    \draw[thick] (\xin,-\H) rectangle (\xout,\H);
    \draw[dashed, gray] (\xin,0) -- (\xout,0);
    \draw[thick, fill=gray!45] (0,0) circle (0.5);
    \draw[->] (0,0) -- (1.6,0) node[below] {$x$};
    \draw[->] (0,0) -- (0,1.6) node[left] {$y$};
    \draw[->] (1.9,1.4) node[right] {$\Gamma_\text{c}$: $u=v=0$} -- (0.38,0.33);
    \foreach \y in {-4,-2,2,4}
        \draw[->, thick] (\xin-1.3,\y) -- (\xin-0.1,\y);
    \node[left, align=center] at (\xin-0.15,0) {$u=1$\\$v=0$};
    \node[right] at (\xin,4.3) {$\Gamma_\text{in}$};
    \node[above] at (4.5,\H) {$\Gamma_\text{lat}$: $v=0$, $\partial u/\partial y=0$};
    \node[below] at (4.5,-\H) {$\Gamma_\text{lat}$: $v=0$, $\partial u/\partial y=0$};
    \node[left] at (\xout,4.3) {$\Gamma_\text{out}$};
    \node[right, align=left] at (\xout+0.2,2.6) {no\\condition};
    \node[gray!40!black] at (5.3,3.2) {$t=0$: $u=3/2$, $v=0$};
    \node[gray!40!black] at (5.3,-3.2) {$t=0$: $u=1/2$, $v=0$};
    \node at (-2.8,-3.2) {$\Omega$};
    \draw (\xin,-\H) -- (\xin,-\H-0.9) node[below] {$-5$};
    \draw (0,-\H) -- (0,-\H-0.9) node[below] {$0$};
    \draw (\xout,-\H) -- (\xout,-\H-0.9) node[below] {$9.3$};
    \draw (\xout,\H) -- (\xout+0.4,\H) node[right] {$5$};
    \draw (\xout,-\H) -- (\xout+0.4,-\H) node[right] {$-5$};
\end{tikzpicture}
\caption{Computational domain $\Omega$, to scale, boundary conditions and initial condition~\eqref{eq:ic}. The cylinder of unit diameter is centered at the origin, on the wake centerline $y=0$ (dashed). $\Gamma_\text{in}$, $\Gamma_\text{out}$, $\Gamma_\text{lat}$ and $\Gamma_\text{c}$ denote the inlet, the outlet, the lateral boundaries and the cylinder wall.}
\label{fig:domain}
\end{figure}

with $\bfv=(u,v)^\intercal$, $p$ the pressure scaled by $\rho U_\infty^2$, and $\mathit{Re}=U_\infty D/\nu=100$. The cylinder of unit diameter is centered at the origin of the rectangle $[-5,\,9.3]\times[-5,\,5]$, which gives a blockage ratio of $0.1$ (Fig.~\ref{fig:domain}). The boundary conditions are $u=1$, $v=0$ at the inlet, no slip on the cylinder, and $v=0$, $\partial u/\partial y=0$ on the lateral boundaries $y=\pm5$. No condition is imposed at the outlet; the influence of this choice is left for future work. The domain and the boundary conditions are symmetric about $y=0$, so that a symmetric initial condition would lead to a symmetric solution. The symmetry is broken by the initial condition
\begin{equation}\label{eq:ic}
    u(0,x,y)=\begin{cases}3/2 & \text{if } y\geq0,\\ 1/2 & \text{if } y<0,\end{cases}\qquad v(0,x,y)=0,
\end{equation}
two uniform streams separated by a shear at $y=0$. It is divergence-free and given in closed form, and its flow rate through any section $x=\text{const.}$ equals that imposed at the inlet. It is obtained by adding $1/2$ to the initial condition used to trigger shedding with the LS-STAG cut-cell method~\cite{cheny2010ls}. It satisfies neither the no-slip condition on the cylinder nor the inlet condition. No other information on the flow is given to the network.

\subsection{PINN over successive time windows}
The time interval is divided into windows $[t_k,t_{k+1}]$ of length $T_w=0.4$, about $7\%$ of the shedding period, with $t_k=k\,T_w$. A network $\mathcal{N}_k:(x,y,t)\mapsto(u,v,p)$ is trained on each window, the windows being treated one after the other, and $\mathcal{N}_{k+1}$ is initialized with the parameters of $\mathcal{N}_k$. Within a window, time is a continuous input of the network and $\partial_t\bfv$ is computed by automatic differentiation: no temporal discretization is introduced. The loss of $\mathcal{N}_{k+1}$ reads
\begin{equation}\label{eq:loss}
    \mathcal{L}=\mathcal{L}_\text{res}+\mathcal{L}_\text{bc}+\omega_\text{ic}\,\mathcal{L}_\text{ic},
\end{equation}
where $\mathcal{L}_\text{res}$ is the mean square of the residuals of~\eqref{eq:NS} at collocation points drawn at random from the uniform distribution on $[t_k,t_{k+1}]\times\Omega$ and redrawn at each iteration (random sampling with resampling~\cite{wu2023comprehensive}), $\mathcal{L}_\text{bc}$ that of the boundary conditions, and $\mathcal{L}_\text{ic}$ the mean square of $\bfv_{\mathcal{N}_{k+1}}(t_k)-\bfv_{\mathcal{N}_k}(t_k)$ at points drawn in $\Omega$, with $\omega_\text{ic}=10$, the initial condition of the first window being~\eqref{eq:ic}. This is the $\bfv$-junction. The $(\bfv,p)$-junction adds the pressure to $\mathcal{L}_\text{ic}$, and the consistent junction of Section~\ref{sec:remedy} adds, in addition, a term $\mathcal{L}_{\text{div},t}$ to~\eqref{eq:loss}. A control adds this term to the $\bfv$-junction without the pressure. The four configurations use the same network and training parameters, summarized in Table~\ref{tab:setup}. Other techniques improve the accuracy of PINNs, such as the adaptive refinement of the collocation points~\cite{wu2023comprehensive} or the dynamic weighting of the terms of the loss~\cite{xiang2021self}. None of them is used here: the four configurations differ only by the initial condition of each window and by the term $\mathcal{L}_{\text{div},t}$, so as to isolate the effect of the junction.

\begin{table}[ht!]
\begin{center}
\caption{Junctions between windows and common parameters. All runs use $T_w=0.4$, GELU activations, Fourier features of the spatial coordinates and of time~\cite{tancik2020fourier,wang2023expert}, and the parameters of $\mathcal{N}_k$ to initialize $\mathcal{N}_{k+1}$.}
\label{tab:setup}
\small\setlength{\tabcolsep}{4pt}
\begin{tabular}{l l l}
\hline
Junction & Initial condition of $\mathcal{N}_{k+1}$ & Hidden constraint $\mathcal{L}_{\text{div},t}$ \\
\hline
$\bfv$-junction & $\bfv$ & no \\
$\bfv$-junction $+\,\mathcal{L}_{\text{div},t}$ (control) & $\bfv$ & yes, weight 1 \\
$(\bfv,p)$-junction & $\bfv$ and $p$ & no \\
Consistent junction & $\bfv$ and $p$ & yes, weight 1 \\
\hline
Hidden layers & \multicolumn{2}{l}{$3\times193$} \\
Spatial Fourier features & \multicolumn{2}{l}{128, $\sigma=2.5$} \\
Time Fourier features & \multicolumn{2}{l}{128, $\sigma_t=0.25$} \\
Optimizer & \multicolumn{2}{l}{Adam, learning rate $10^{-4}$} \\
Iterations per window & \multicolumn{2}{l}{$10^4$} \\
Collocation points per iter. & \multicolumn{2}{l}{1024, uniform random, redrawn} \\
\hline
\end{tabular}
\end{center}
\end{table}

\subsection{\label{sec:qoi}Quantities of interest}
The drag and lift coefficients $C_D$ and $C_L$ are obtained by integrating the pressure and the viscous stress given by the network on the cylinder, with the composite midpoint rule on 200 arcs. The Strouhal number $\mathit{St}$ is obtained from the upward zero crossings of $C_L-\overline{C}_L$, located by linear interpolation. Each network is evaluated at 21 equally spaced instants of its window, both ends included, so that at a junction $t_k$ both $\mathcal{N}_k$ and $\mathcal{N}_{k+1}$ are evaluated; the statistics use the value of $\mathcal{N}_k$. The pressure is defined up to an additive constant, which may differ from one window to the next; at a probe we therefore report $p-p_\text{ref}$, with $p_\text{ref}$ the pressure at $(-4.49,\,0.01)$, upstream of the cylinder. The probes A, B and C are located at $(0,\,0.8)$, above the cylinder, $(1,\,0)$, in the recirculation region, and $(4,\,0)$, on the wake centerline, to within $0.015$: each probe is the node of the reference mesh nearest to these points, and both solutions are evaluated there. With $[q]_k=q_{\mathcal{N}_{k+1}}(t_k)-q_{\mathcal{N}_k}(t_k)$ the jump of a quantity $q$ at $t_k$, we define the jump ratio and the slope-jump ratio
\begin{equation}\label{eq:jumpratio}
    \tau_k(q)=\frac{\bigl|[q]_k\bigr|}{\Delta q_\text{ref}},\qquad
    \kappa_k(q)=\frac{\bigl|[\partial_t q]_k\bigr|}{\max_t|\partial_t q_\text{ref}|},
\end{equation}
where $\Delta q_\text{ref}$ is the half peak-to-peak amplitude of $q$ and $\max_t|\partial_t q_\text{ref}|$ the maximum magnitude of its time derivative, both taken from the periodic regime of the reference solution described below, and $[\partial_t q]_k$ is computed by automatic differentiation of $\mathcal{N}_k$ and $\mathcal{N}_{k+1}$ at $t_k$. The jump ratio $\tau_k$ is the jump as a fraction of the reference amplitude, and the slope-jump ratio $\kappa_k$ is the jump of the slope as a fraction of the maximum reference slope; it is only meaningful where $q$ is continuous. Both scales are taken from the reference solution rather than from the PINN signal, so that they are the same for the four configurations and are not affected by the errors of each run on the amplitude or on the slope of $q$. We report the medians of $\tau_k$ and $\kappa_k$ over the junctions. All the networks are evaluated in double precision.

The reference solution is computed with OpenFOAM (\texttt{icoFoam}, second order in space and time) on the same domain, from rest, on a mesh of $139\,056$ cells; the statistics are taken over $t\in[130,200]$. On a mesh refined by a factor $1.5$, $\overline{C}_D$ and $\mathit{St}$ change by $0.2\%$ and $0.1\%$, and $C_L'$ by $1.2\%$. This solution uses a fixed pressure and a zero velocity gradient at the outlet.

\section{\label{results}Results}

\subsection{\label{sec:base}Onset of shedding}
\begin{figure}[ht!]
\centering
\includegraphics[width=\textwidth]{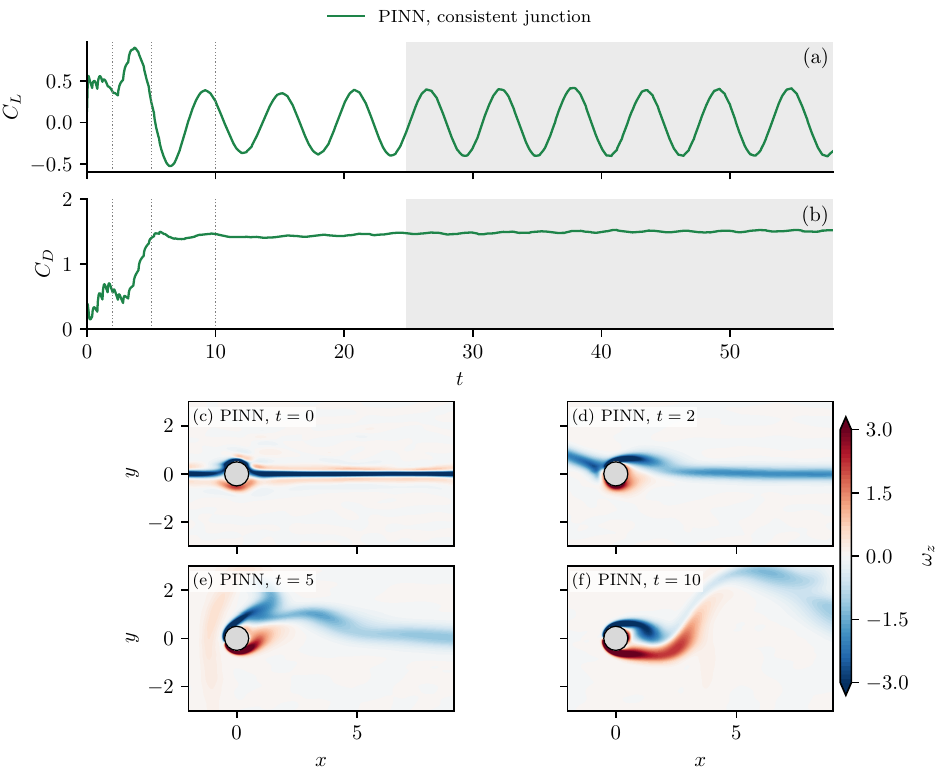}
\caption{Onset of vortex shedding computed by the PINN with the consistent junction, from the initial condition~\eqref{eq:ic}. (a) Lift and (b) drag coefficients; the dotted lines mark the instants shown below and the shaded interval is that of the statistics. (c)--(f) Vorticity of the PINN at $t=0$, $2$, $5$ and $10$. The reference solution, computed from rest, has a different transient and is not shown.}
\label{fig:onset}
\end{figure}

Figure~\ref{fig:onset} shows the force coefficients obtained with the consistent junction (Section~\ref{sec:remedy}) over $t\in[0,58]$, i.e., 145 windows, and the vorticity at four instants. The network replaces the discontinuity of~\eqref{eq:ic} at $y=0$ by a thin shear layer. At $t=2$, the wake behind the cylinder is already asymmetric, at $t=5$ a vortex rolls up in the near wake, and at $t=10$ vortices of opposite sign are shed alternately into the wake. No steady symmetric recirculation region forms before the shedding. From $t\approx8$, the half peak-to-peak amplitude of $C_L$ over each period lies between $0.37$ and $0.41$. The four configurations of Table~\ref{tab:setup} develop shedding from the same initial condition, and their Strouhal numbers differ from the reference value by less than $1.5\%$ (Table~\ref{tab:stats}). All the statistics and junction ratios below are computed over $t\in[24.8,58]$, i.e., about five shedding periods and $83$ junctions. The interval stops at $t=58$ because, beyond it, the mean drag coefficient of the consistent junction drifts (Section~\ref{conclusion}).

\begin{table}[ht!]
\begin{center}
\caption{Statistics of the periodic regime, and medians of the jump ratio $\tau_k$ and of the slope-jump ratio $\kappa_k$~\eqref{eq:jumpratio} at the junctions, both normalized by the reference solution, for $t\in[24.8,58]$ (83 junctions); reference solution: $t\in[130,200]$. $S_\text{lat}$: maximum of $|\int\nabla\cdot\bfv\,\mathrm{d}x\,\mathrm{d}y|$ over the bands $|y|>3$. The probes are those of Fig.~\ref{fig:probesbase}. $\kappa_k$ is not given for the $\bfv$-junction and its control, whose $C_L$ and pressure are discontinuous.}
\label{tab:stats}
\scriptsize\setlength{\tabcolsep}{3pt}
\begin{tabular}{l c c c c c}
\hline
 & Reference & $\bfv$ & \begin{tabular}{@{}c@{}}$\bfv$\\$+\,\mathcal{L}_{\text{div},t}$\end{tabular} & $(\bfv,p)$ & Consistent \\
\hline
IC of $\mathcal{N}_{k+1}$ & -- & $\bfv$ & $\bfv$ & $\bfv$, $p$ & $\bfv$, $p$ \\
$\mathcal{L}_{\text{div},t}$ & -- & no & yes & no & yes \\
\hline
$\mathit{St}$ & 0.178 & 0.179 & 0.178 & 0.176 & 0.178 \\
$\overline{C}_D$ & 1.492 & 1.487 & 1.478 & 1.499 & 1.495 \\
$C_L'$ & 0.261 & 0.273 & 0.260 & 0.288 & 0.284 \\
$\Delta C_L$ & 0.370 & 0.415 & 0.387 & 0.491 & 0.412 \\
$S_\text{lat}$ & -- & 0.04 & 0.03 & 0.26 & 0.07 \\
\hline
$\tau_k(C_L)$ & -- & 0.19 & 0.11 & 0.007 & 0.009 \\
$\tau_k(C_D)$ & -- & 0.93 & 0.81 & 0.30 & 0.16 \\
$\tau_k(p-p_\text{ref})$, A/B/C & -- & 0.27/0.16/0.21 & 0.10/0.13/0.08 & 0.01/0.03/0.01 & 0.006/0.020/0.009 \\
$\tau_k(u)$, $\tau_k(v)$, A, B, C & -- & 0.001--0.03 & 0.001--0.03 & 0.002--0.05 & 0.002--0.04 \\
\hline
$\kappa_k(C_L)$ & -- & -- & -- & 1.2 & 0.24 \\
$\kappa_k(p-p_\text{ref})$, A/B/C & -- & -- & -- & 14/17/9.2 & 0.28/0.28/0.18 \\
\hline
\end{tabular}
\end{center}
\end{table}

\subsection{\label{sec:junctions}A staircase lift coefficient}
\begin{figure}[ht!]
\centering
\includegraphics[width=\textwidth]{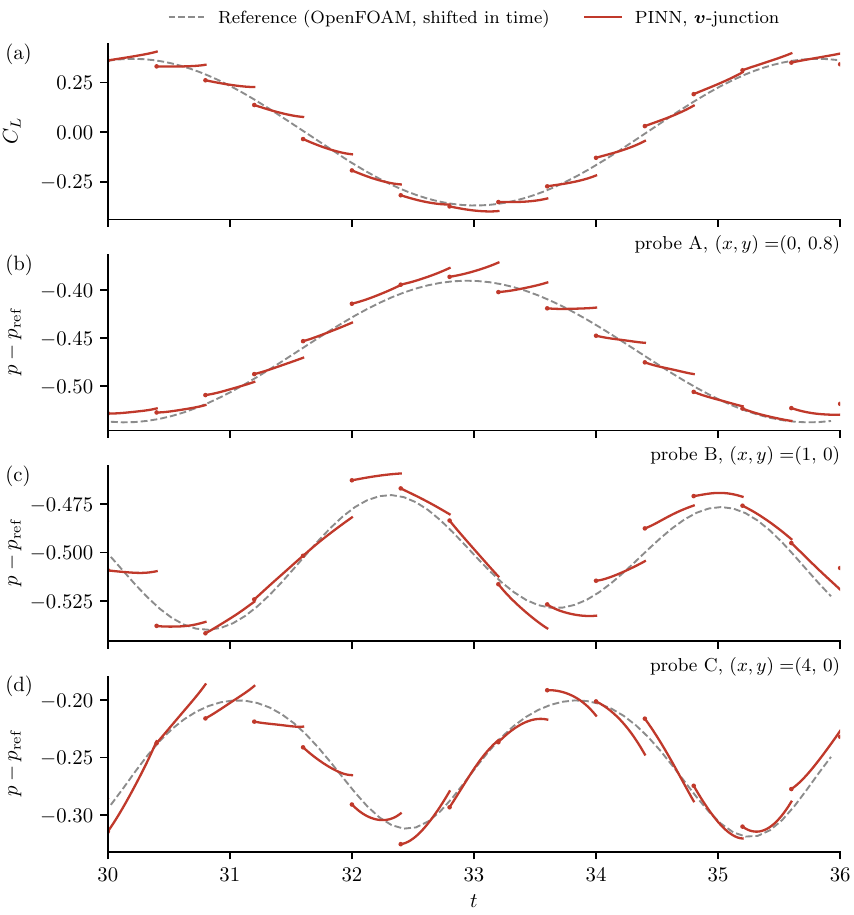}
\caption{PINN with the $\bfv$-junction (solid lines) and reference solution (OpenFOAM, dashed lines): (a) lift coefficient and (b)--(d) pressure $p-p_\text{ref}$ at the probes A, B and C, over $t\in[30,36]$. Each window of the PINN is plotted separately, its first instant being marked by a dot. The reference solution is shifted in time to match the phase of $C_L$.}
\label{fig:probesbase}
\end{figure}

With the $\bfv$-junction, the statistics agree with the reference solution: $\mathit{St}$ and $\overline{C}_D$ are within $0.5\%$ of the reference values (Table~\ref{tab:stats}). This agreement hides a defect, visible when each window is plotted separately (Fig.~\ref{fig:probesbase}a): $C_L$ jumps at every junction, and $53\%$ of its total variation over $t\in[24.8,58]$ takes place through these jumps. The lift coefficient is a staircase. The rms value of the jumps $[C_L]_k$ is $0.074$, i.e., $27\%$ of $C_L'$, and the median jump is $19\%$ of the reference amplitude $\Delta C_L$ ($\tau_k=0.19$). The amplitude $\Delta C_L$ is overestimated by $12\%$. The drag coefficient jumps as well, by $0.016$ in rms value ($\tau_k=0.93$). Khan et al.~\cite{khan2026physics} observe the same kind of jump of the forces at the junction between two windows.

To identify the discontinuous field, Fig.~\ref{fig:probesbase}b--d shows $p-p_\text{ref}$ at the three probes A, B and C (Section~\ref{sec:qoi}). The pressure is discontinuous at every junction, the median jump being $27\%$, $16\%$ and $21\%$ of the reference amplitude at A, B and C. Between two junctions, the pressure follows the variation of the reference solution. The velocity components at the same probes are continuous, with jump ratios between $0.001$ and $0.03$.

This discontinuity follows from the structure of~\eqref{eq:NS}. The pressure has no evolution equation: it is an algebraic variable, determined at each instant by the velocity through the constraint~\eqref{eq:cont}, once this constraint is differentiated in time. This is the structure of a differential-algebraic system of index~2, which the spatial discretizations of~\eqref{eq:NS} are~\cite{hairer1989,wan2019accuracy}; the PINN solves~\eqref{eq:NS} without spatial discretization, and the argument below is made on the continuous equations. The initial condition of $\mathcal{N}_{k+1}$ imposes its velocity at $t_k$, but not its pressure. In the exact solution, the pressure is fixed, up to a constant, by the constraint
\begin{equation}\label{eq:hidden}
    \nabla\cdot\partial_t\bfv=0,
\end{equation}
obtained by differentiating~\eqref{eq:cont} in time, which we call the hidden constraint; through~\eqref{eq:mom}, it is, for a divergence-free velocity, a Poisson equation for the pressure. As for an index-2 system~\cite{hairer1989,wan2019accuracy}, an initial condition is consistent only if it satisfies this hidden constraint. In the limit of a vanishing loss, the constraint~\eqref{eq:cont}, enforced over the whole window, would imply~\eqref{eq:hidden}. For index-2 systems, time integrators lose accuracy on the algebraic variable, here the pressure: its order of convergence is lower than that of the differential variable, here the velocity~\cite{hairer1989,wan2019accuracy}. By analogy with this loss of accuracy, we interpret the pressure jumps at the junctions as follows: the loss controls $\nabla\cdot\bfv$ only in mean square over a short window, and therefore determines the pressure at the junction only weakly.

\subsection{\label{sec:remedy}Remedy: the consistent junction}
The remedy, the consistent junction, has two ingredients, which we separate with two controls that contain only one of them: the $(\bfv,p)$-junction and the $\bfv$-junction with $\mathcal{L}_{\text{div},t}$. First, the initial condition of $\mathcal{N}_{k+1}$ imposes both the velocity and the pressure of $\mathcal{N}_k$ at $t_k$. Second, the hidden constraint~\eqref{eq:hidden} is added to the loss~\eqref{eq:loss} through the term
\begin{equation}\label{eq:lossdivt}
    \mathcal{L}_{\text{div},t}=\frac{1}{N_r}\sum_{i=1}^{N_r}\Bigl|\partial_t\bigl(\nabla\cdot\bfv\bigr)(t_i,x_i,y_i)\Bigr|^2,
\end{equation}
evaluated at the $N_r$ collocation points of the residuals, with a weight of 1. The constraint~\eqref{eq:cont} and its time derivative~\eqref{eq:hidden} are thus both enforced; replacing the former by the latter would only preserve $\nabla\cdot\bfv=0$ if the initial condition satisfied it exactly~\cite{wan2019accuracy}. When the residuals of~\eqref{eq:mom} and~\eqref{eq:cont} vanish, $\mathcal{L}_{\text{div},t}=0$ is equivalent to the pressure Poisson equation $\nabla^2p+\nabla\bfv:(\nabla\bfv)^\intercal=0$, whose residual Hu and Senocak~\cite{hu2025unsupervised} minimize, in discrete time. The term itself is therefore not new; its role at the junctions between windows is. The $(\bfv,p)$-junction uses the pressure in the initial condition but not $\mathcal{L}_{\text{div},t}$, and the control $\bfv+\mathcal{L}_{\text{div},t}$ uses $\mathcal{L}_{\text{div},t}$ but not the pressure (Table~\ref{tab:setup}).

\begin{figure}[ht!]
\centering
\includegraphics[width=\textwidth]{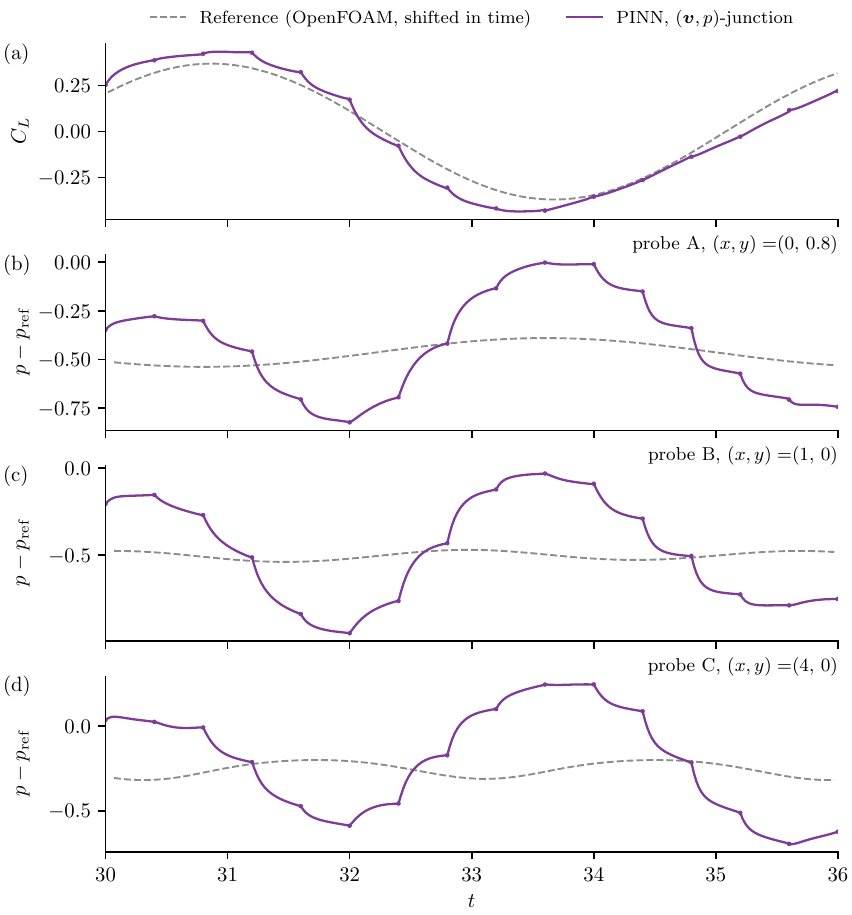}
\caption{PINN with the $(\bfv,p)$-junction, i.e., with the pressure in the initial condition of each window but without $\mathcal{L}_{\text{div},t}$ (solid lines), and reference solution (OpenFOAM, dashed lines): same quantities, probes and time interval as in Fig.~\ref{fig:probesbase}.}
\label{fig:probescontrol}
\end{figure}

Imposing the pressure makes the solution continuous (Fig.~\ref{fig:probescontrol}): with the $(\bfv,p)$-junction, the median jump ratio of $C_L$ falls from $0.19$ to $0.007$, and those of $p-p_\text{ref}$ at the three probes lie between $0.01$ and $0.03$ (Table~\ref{tab:stats}). Continuity is, however, not sufficient. The slope of the signals jumps at each junction: the median jump of the slope is $1.2$ times the maximum slope of the reference solution for $C_L$, and between $9$ and $17$ times for the pressure at the probes. Between two junctions, the pressure at the probes departs from the reference solution, with an amplitude several times larger (Fig.~\ref{fig:probescontrol}b--d). The lift amplitude $\Delta C_L$ is $33\%$ too large, and the Strouhal number, $0.176$, is $1.4\%$ below the reference value. Without~\eqref{eq:hidden}, the time derivative of the velocity associated with the transmitted pressure through~\eqref{eq:mom} need not be divergence-free, and we observe spurious sources and sinks of mass: the integral of $\nabla\cdot\bfv$ over the bands $|y|>3$ reaches $0.26$, i.e., $2.6\%$ of the flow rate through the domain, against $0.04$ with the $\bfv$-junction.

The hidden constraint alone does not remove the jumps either. With the control $\bfv+\mathcal{L}_{\text{div},t}$, the median jump ratio of $C_L$ falls from $0.19$ to $0.11$, and those of $p-p_\text{ref}$ at A, B and C from $0.27$, $0.16$ and $0.21$ to $0.10$, $0.13$ and $0.08$ (Table~\ref{tab:stats}), i.e., by factors between $1.3$ and $2.7$: $C_L$ and the pressure remain discontinuous. The Strouhal number is within $0.5\%$ of the reference value, $\overline{C}_D=1.478$ is $0.9\%$ below it, and the mass source is $0.03$. Neither ingredient is thus sufficient alone: the pressure makes the solution continuous but not consistent, and the hidden constraint reduces the jumps without removing them.

\begin{figure}[ht!]
\centering
\includegraphics[width=\textwidth]{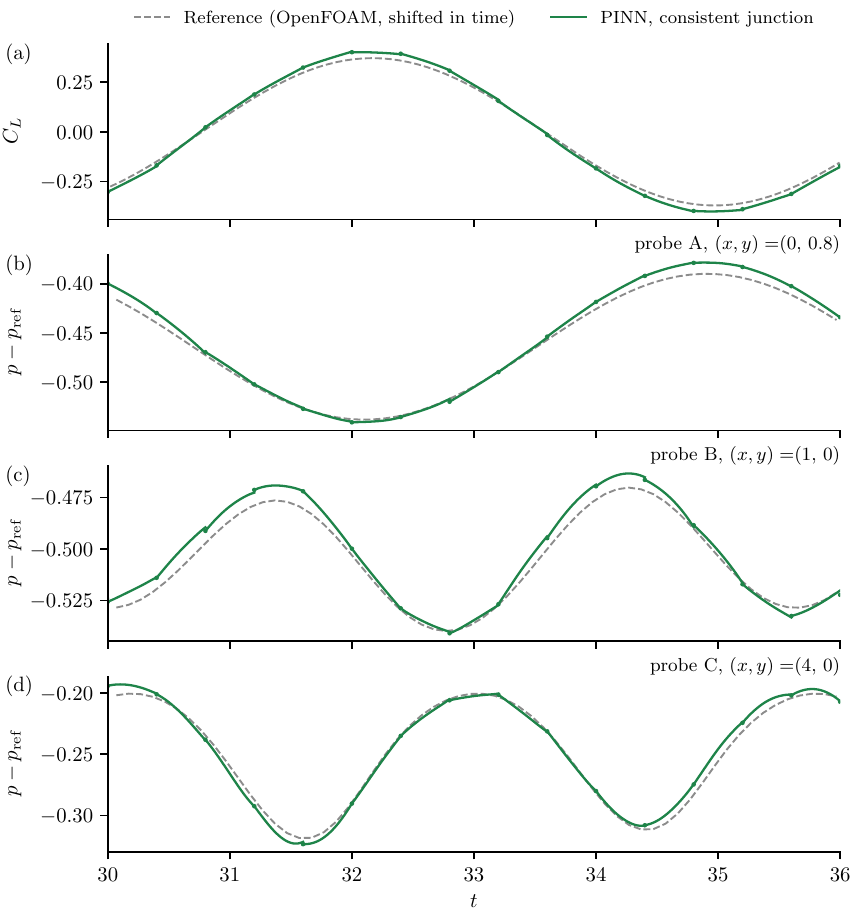}
\caption{PINN with the consistent junction, i.e., with the pressure in the initial condition of each window and $\mathcal{L}_{\text{div},t}$ in the loss (solid lines), and reference solution (OpenFOAM, dashed lines): same quantities, probes and time interval as in Fig.~\ref{fig:probesbase}.}
\label{fig:probesremedy}
\end{figure}

The consistent junction, which adds $\mathcal{L}_{\text{div},t}$, removes these defects (Fig.~\ref{fig:probesremedy}). The lift coefficient and the pressure remain continuous, with a median jump ratio of $0.009$ for $C_L$, $20$ times smaller than with the $\bfv$-junction, and between $0.006$ and $0.02$ for the pressure. The slope-jump ratios fall to $0.24$ for $C_L$ and between $0.18$ and $0.28$ for the pressure, i.e., $5$ times smaller for $C_L$ and $50$ to $60$ times smaller for the pressure than with the $(\bfv,p)$-junction, and the mass source falls to $0.07$. Between two junctions, the pressure at the probes follows the reference solution. The mean drag coefficient and the Strouhal number are within $0.5\%$ of the reference values. The lift amplitude is less accurate: $C_L'=0.284$ and $\Delta C_L=0.412$ exceed the reference values by $9\%$ and $11\%$. The mean drag coefficient increases slowly over the interval, from $1.472$ over its first period to $1.509$ over its last one. The pressure imposed at the junctions thus ensures the continuity of the solution, and the hidden constraint ensures that this pressure is consistent with the velocity.

\begin{figure}[ht!]
\centering
\includegraphics[width=\textwidth]{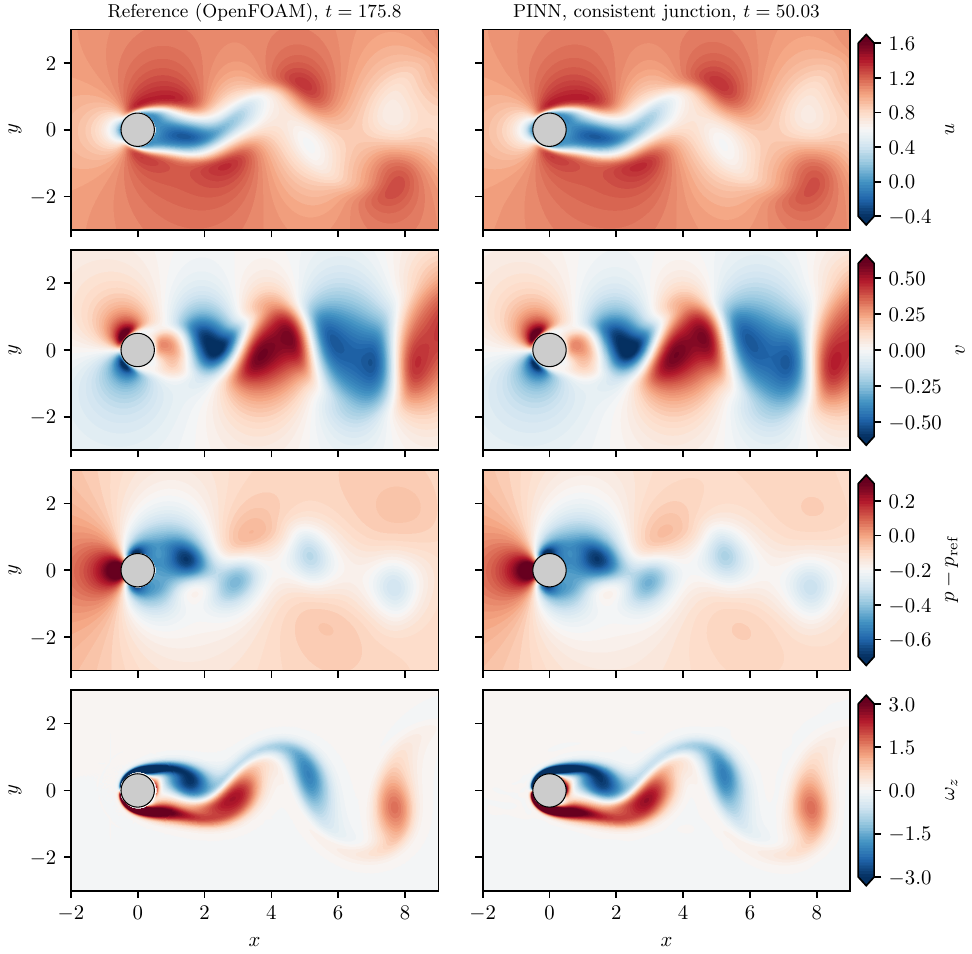}
\caption{Instantaneous fields $u$, $v$, $p-p_\text{ref}$ and $\omega_z=\partial_xv-\partial_yu$ at the same phase of the shedding cycle: reference solution (OpenFOAM, left) and PINN with the consistent junction (right). The reference solution is shifted in time so that its lift coefficient matches that of the PINN (least squares over $t\in[44,50]$); $t=175.8$ for the reference corresponds to $t=50.03$ for the PINN. The vorticity of the reference solution is computed by finite differences and is not shown within $0.03$ of the wall.}
\label{fig:fields}
\end{figure}

At the same phase of the shedding cycle, the instantaneous fields of the consistent junction are close to those of the reference solution (Fig.~\ref{fig:fields}): over the region shown, the relative $L^2$ differences are $1.5\%$ for $u$, $7\%$ for $p-p_\text{ref}$, and $12\%$ and $10\%$ for $v$ and $\omega_z$.

\section{\label{conclusion}Conclusion}
A continuous-time PINN trained over successive time windows develops vortex shedding past a circular cylinder at $\mathit{Re}=100$, in a domain symmetric about the wake centerline, without reference data and without temporal discretization. The symmetry is broken by an analytical, divergence-free initial condition, and the Strouhal number and the mean drag coefficient agree with an OpenFOAM solution on the same domain. With the $\bfv$-junction, the lift coefficient is a staircase, because the pressure jumps at the junctions between windows while the velocity is continuous. The spatial discretizations of the incompressible Navier--Stokes equations are differential-algebraic systems of index~2~\cite{hairer1989,wan2019accuracy}: a consistent initial condition must satisfy $\nabla\cdot\bfv=0$ for the velocity, and the hidden constraint $\nabla\cdot\partial_t\bfv=0$, with $\partial_t\bfv$ given by the momentum equation, then determines the pressure. The $\bfv$-junction fixes the velocity only, and, by analogy with the loss of accuracy of time integrators on the algebraic variable of index-2 systems, we interpret the jumps as a weak determination of the pressure at the junction by the loss; the continuity of the velocity is not sufficient. Imposing the pressure in the initial condition of each window makes the solution continuous, but not consistent: without the hidden constraint in the loss, the slope of the signals jumps at the junctions, and spurious mass sources appear. The hidden constraint without the pressure reduces the jumps by factors between $1.3$ and $2.7$ only. With both the pressure and the hidden constraint, i.e., with the consistent junction, the lift and the pressure are continuous, the median jump of the lift is $20$ times smaller than with the $\bfv$-junction, and the mean drag coefficient and the Strouhal number agree with the reference solution over five shedding periods.

The main open question is the long-time behavior of the consistent junction. Beyond $t\approx58$, its mean drag coefficient increases, from $1.51$ to about $1.6$ within two shedding periods. Over $t\in[25,200]$, $\overline{C}_D$ is $1.500$ with the $\bfv$-junction, $1.486$ with the control $\bfv+\mathcal{L}_{\text{div},t}$ and $1.587$ with the consistent junction, for a reference value of $1.492$; with $4\times193$ hidden layers, these values are $1.496$, $1.490$ and $1.575$. The $(\bfv,p)$-junction keeps $\overline{C}_D=1.486$, but its lift amplitude $\Delta C_L$ reaches $1.70$, against $0.370$ for the reference solution. The degradation beyond $t\approx58$ thus appears only with the junctions that impose the pressure. The pressure transmitted at each junction carries the error of the previous window. The observed degradation is consistent with an accumulation of this error from one window to the next, but does not prove it. The onset of shedding from rest, i.e., without the asymmetry of the initial condition, also remains to be studied. These questions, together with the influence of the amplitude of the initial asymmetry and of the window length, are the subject of a parametric study in progress.

\section*{Acknowledgments}
This work benefited from the support of engineers in artificial intelligence within a program funded by the French government through the France 2030 investment plan managed by the National Research Agency (ANR), under reference ANR-24-RIIA-0001.

\bibliographystyle{elsarticle-num}
\bibliography{biblio}

\end{document}